\documentclass[prb,preprint]{revtex4-1} 
\usepackage{amsmath}  
\usepackage{amsfonts} 
\usepackage{graphicx}
\usepackage{subfigure}
\usepackage{dcolumn}
\usepackage{bm}
\usepackage{url}

\begin{document}

\title{The Motion Of A Spring Released From Uniform Circular Motion}

\author{Thomas Dooling}%
 \affiliation{Department of Physics, University of North Carolina at Pembroke, Pembroke, NC 28372.}
\author{Jeffrey Regester }
 \affiliation{Greensboro Day School, Greensboro, NC 27410.}
\author{Matthew Carnaghi}%
 \affiliation{Department of Physics, High Point University, High Point, NC 27268.}
\author{Aaron Titus}%
 \affiliation{Department of Physics, High Point University, High Point, NC 27268.}

\date{\today}

\begin{abstract}
A weak spring is connected at one end to a rotor turning at constant angular velocity. The spring extends to a stretched length as determined by the spring mass, rest length, spring constant, rotor radius and rotor angular velocity. When released from the rotor, the inner end of the spring pulls away as expected, causing a wave to travel down the spring as it collapses. During this time interval, the outer end of the spring continues to move along its original circular path in uniform circular motion, as if the spring were still connected to the rotor.  This is analogous to the effect of a hanging Slinky released from rest whose bottom end remains at a fixed position above the ground until a wave from the top of the Slinky reaches the bottom of the Slinky. Values from a numerical model and measurements from video analysis show that upon release the inner end travels along a circle of similar radius as the outer end. The effect appears as a series of alternating semi-circles. In addition, the simulation and data agree that (1) the spring extension and drag angle increase with the angular velocity of the rotor; (2) the droop angle decreases with angular velocity of the rotor; (3) the collapse time and bend angle of the collapsing spring are independent of the angular velocity.

\end{abstract}

\maketitle

\section{\label{sec:Experiment}Experiment}
A long, weak spring (such as a Slinky) hanging vertically, held at one end reaches an equilibrium length under the pull of gravity, if given sufficient time for oscillations to dampen. When the top end is released, the stationary bottom end of the spring does not react until sufficient time has elapsed for a wave to propagate from the top end to the bottom. Calkin \cite{calkin} and Cross and Wheatland \cite{rccross:ajp12} develop the theory behind this phenomenon.  Seen in person or especially in high-speed video\cite{macisaac:tpt13}$^,$\cite{veritasium:awesomeHDslinky} the phenomenon is striking, even if expected.

Here we extend this experiment into the realm of circular motion. If a spring is whirled in a circle with one end free, the free end  extends outward due to its own inertia or, expressed in terms of a rotating reference frame, the action of centrifugal force. Given sufficient time for initial oscillations to dampen, the free end travels in uniform circular motion. If the inner end of the spring is released, we expect the free end to continue in uniform circular motion as before, until such time as a wave propagates from the inner end to the outer. 

\begin{figure}[h!]
	\centering
	\includegraphics[height=2.5in]{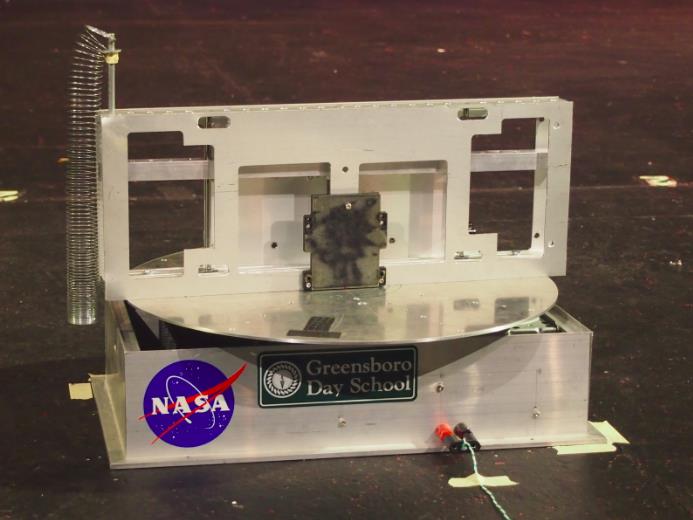}
	\caption{A photo of the apparatus, stationary, with the thin Slinky hanging from the raised pin.}
	\label{apparatus}
\end{figure}

To test this hypothesis, we use a custom-built turntable shown in Fig. \ref{apparatus}  that has a hold-and-release mechanism. The turntable has a maximum angular speed, unloaded, of 270 revolutions per minute. A tall rectangular box sitting atop the turntable contains a horizontal bar into which 0.25 inch pins may be affixed, at distances from 2 cm to 24 cm from the axis, in 2-cm increments, on both sides of the axis. For this experiment only one pin was used, positioned 24 cm from the turntable axis, onto which one end of the subject spring is hooked. When the bar is manually raised, a strong internal spring is compressed and the bar is latched. When the turntable is rotating, a signal from an infrared remote control triggers a servo which trips the latch. The bar drops, retracting the pin and releasing the inner end of the whirling spring. We generally refer to the apparatus holding the spring as the rotor. An excellent video \cite{regester:spinningsprings} demonstrating the device will help the reader to better visualize the process.

Experimental data were collected in the form of high-speed (300 fps and 600 fps) video recorded from a position 8.4 m above the floor on which the turntable was sitting. Videos were analyzed using the free, open-source video analysis software \emph{Tracker}\cite{trackerweb}$^,$\cite{tracker}. The inner end of the spring is held near the top of the pin,  41.8 cm above the floor, so the rotating spring can droop due to gravity up to 41.8 cm from the horizontal plane without the free end touching the floor.  For low-stiffness springs like a Slinky, the apparatus is set on a table to allow a larger droop of the spring without touching the floor.


Two springs are used for this experiment. One is a standard PASCO spring with a spring constant of 3.28 N/m. The second is a thin Slinky chosen for its low spring constant of 0.38 N/m.

\section{\label{sec:Theory}Quasi-Static Theory}

\begin{figure}[h!]
	\centering
	\includegraphics[height=0.75in]{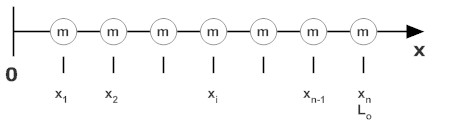}
	\caption{The coordinate system of the spring at rest.}
	\label{fig:restspring}
\end{figure}

The motion of a spinning spring in the absence of air resistance and gravity will first be solved analytically. The result is needed for the simulation discussed later. 
The spring will be modeled as a series of point masses connected by short massless springs (Fig \ref{fig:restspring}). The spring has a rest length $L_o$, and each point mass is separated at regular intervals. A coordinate system placed along the spring is labeled $x$ and each mass is located at positions, $x_1, x_2, ...$ and so on up to the final mass at $x_n$. For an unstretched spring, the separation between any two adjacent masses, $x_i$-$x_{i-1}$, is constant and is labeled $\Delta x$.

\begin{equation}
\Delta x=L_o/n
\label{deltan}
\end{equation}

Each spring between adjacent masses will have a spring constant $k$ given by
\begin{equation}
k=n \times K
\label{lilk}
\end{equation}

where $K$ is the spring constant for the whole spring. Each mass has a value of

\begin{equation}
m=M/n
\label{lilm}
\end{equation}

where $M$ is the total mass of the spring.

\begin{figure}[h!]
	\centering
	\includegraphics[height=1.0in]{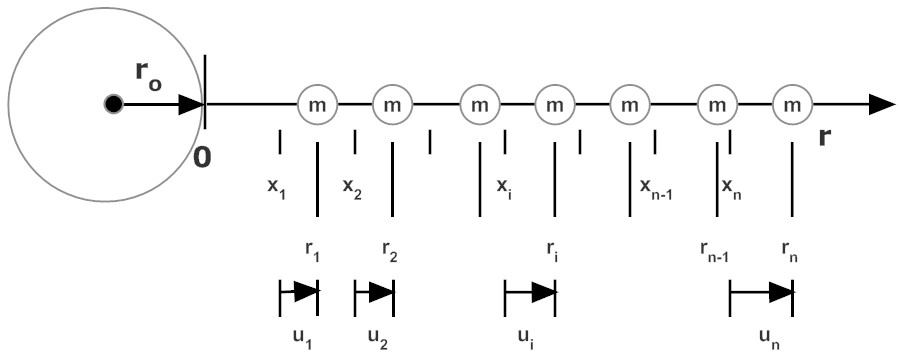}
	\caption{The coordinate system of the stretched spring attached to a rotating rotor of radius $r_o$.}
	\label{fig:stretchspring}
\end{figure}

When the spring is connected to a rotor rotating with an angular velocity $\omega$, it stretches out to a new length. The axis of rotation is labeled $z$. In the rotating reference frame, each mass has a spring force acting to its right and left as well as the fictitious centrifugal force to the right. The masses are located on a coordinate system labeled $r$, and the displacement of a mass from its rest position is $u$ (Fig. \ref{fig:stretchspring}).

Summing the forces on a point mass gives

\begin{equation}
+k(r_{i+1}-r_i)-k(r_i-r_{i-1})=-m\omega^2 r_i
\label{roteqnb}
\end{equation}

Keep in mind that the mass attached to the rotor cannot move before release and is labeled $r_o$.

The equation for the final mass ($i=n$) is different because it is pulled by only one spring and the rest length of the spring is not canceled out by the rest length of a second spring.
\begin{equation}
-k \times (r_n-r_{n-1}-\Delta x)=-m\omega^2 r_n
\label{roteqnrb}
\end{equation}

The above equations can be generalized for a continuous system to give a differential equation for the initial state of the spring in the rotating system. Its solution can be used to find the stretched length of the spring as a function of position and angular velocity $\omega$.

From Fig. \ref{fig:stretchspring}, $r=u+x+r_o$. So, substituting for $r$ in Eq. (\ref{roteqnb}) gives
\begin{equation}
\begin{split}
 +k(u_{i+1}-u_i)-k (u_i-u_{i-1})\\
 +k(x_{i+1}-x_i)-k (x_i-x_{i-1}) =-m\omega^2 r_i
\label{roteqn2}
\end{split}
\end{equation}
But, since all $\Delta x$ values were defined for the rest spring and are the same, they cancel out leaving
\begin{equation}
\begin{split}
 +k(u_{i+1}-u_i)-k(u_i-u_{i-1})=-m\omega^2 (u_i + x_i + r_o)
\label{roteqn3}
\end{split}
\end{equation}

Since, $k=(n)(K)$ and $m=M/n$. Then,
\begin{equation}
\begin{split}
 +(n)(K)(u_{i+1}-u_i)-(n)(K) (u_i-u_{i-1})=\\
-(M/n) \omega^2 (u_i + x_i + r_o)
\label{roteqn4}
\end{split}
\end{equation}

\begin{equation}
\begin{split}
 (u_{i+1}-u_i)- (u_i-u_{i-1})=-(M/Kn^2) \omega^2 (u_i + x_i + r_o)
\label{roteqn5}
\end{split}
\end{equation}

Since $n=L_o/{\Delta x}$, $n$ can be substituted into the above equation, giving
\begin{equation}
\begin{split}
 (u_{i+1}-u_i)- (u_i-u_{i-1})=-\frac{M\Delta x^2}{KL_o^2} \omega^2 (u_i + x_i + r_o)
\label{roteqn6}
\end{split}
\end{equation}

Dividing by $\Delta x$ twice gives,
\begin{equation}
\begin{split}
\frac{ \frac{(u_{i+1}-u_i)}{\Delta x}- \frac{(u_i-u_{i-1})}{\Delta x}}{\Delta x}=-\frac{M\omega^2}{KL_o^2} (u_i + x_i + r_o)
\label{roteqn7}
\end{split}
\end{equation}

In the limit as $\Delta x$ goes to zero, the left side of the above equation becomes the second derivative of $u$ with respect to x, giving

\begin{equation}
\frac{d^2u}{dx^2}=-\frac{M\omega^2}{KL_o^2}(u + x + r_o)
\label{diffeqn}
\end{equation}

Now, let 
\begin{equation}
\frac{M\omega^2}{KL_o^2}=\alpha^2
\label{alpha-eq}
\end{equation}

Then upon reordering,

\begin{equation}
\frac{d^2u}{dx^2}+\alpha^2 u=-\alpha^2 (x + r_o)
\label{diffeqn2}
\end{equation}

The solution to this equation will give the position of a point on a rotating spring relative to its position when the spring is at rest. Two boundary conditions are needed to solve the equation. The position of the first mass will give the radius of the rotor. Also, the first derivative of $u$ is set equal to zero at the end of the spring since the final mass is infinitesimal and should not cause any stretching of the spring. Using the standard tools for ordinary differential equations yields the following solution.

\begin{equation}
\begin{split}
u=A\cos(\alpha x +\phi)-(x+r_o)\\
A=\frac{r_o}{cos(\phi)}\\
\tan(\phi)= \frac{(-1/(r_o \alpha)-\sin(\alpha L_o))}{\cos(\alpha L_o)}
\label{gensol}
\end{split}
\end{equation}

Since, $r=u+(x+r_o)$, the stretched position of a point mass on the spring is,

\begin{equation}
r=A\cos(\alpha x +\phi)=\frac{r_o}{cos(\phi)}\cos(\alpha x +\phi)
\label{r-quasistatic-eq}
\end{equation}

The inner end of the spring is at the $x=0$ position, and the solution gives the correct position the radius of the rotor.

\begin{equation}
r_{inner\, end}=\frac{r_o}{cos(\phi)}\cos(\phi)=r_o
\end{equation}

The outer end of the spring is at the $x=L_o$ position, so the predicted stretch for the end of the spring is,
\begin{equation}
r_{outer\, end}=\frac{r_o}{cos(\phi)}\cos(\alpha L_o +\phi)
\end{equation}

Finally, if the spring is not spinning, then $\alpha$ is equal to zero. However, naively putting $\alpha=0$ in the solution will give the wrong answer. One must take the limit as $\alpha$ goes to zero, then the outer position will give the rest length of the spring plus the rotor radius.

\section{\label{sec:Simulation}Simulation}
The real spring will be deflected a small amount by a quadratic air resistance term. Also, once the spring is released, the motion will produce coupled sets of equations. For these reasons, the system was numerically simulated and results were compared with experimental data.

\subsection{Initial Condition}

The motion of a spring released from a rotating rotor was simulated using the equations found above. The program, Easy Java Simulations (EJS)\cite{ejs}$^,$\cite{ejsweb}$^,$\cite{ejstpt} was employed to create the simulation. EJS is a Java-based application for numerical modeling using the Open Source Physics\cite{ospweb}$^,$\cite{osp} library. A Runge-Kutta 4 algorithm was used to numerically solve the differential equations. EJS comes with a wide selection of numerical integration algorithms. Runge-Kutta 4  was used in this simulation because it is a well known-algorithm sufficiently sophisticated to yield robust results, yet is among the first numerical integration techniques taught at the undergraduate, and even high-school, level.

Though the stretched length was solved for the continuous case, the simulation is for a discrete set of masses. However, the solution for the continuous case will give initial positions for the masses very close to where they should be. The system is then allowed to relax into its quasi-static state under the effects of gravity and air resistance. If the masses are started too far from their initial positions, the system will become to unstable and the simulation will fail.

\begin{figure}[h!]
	\centering
	\includegraphics[height=2in]{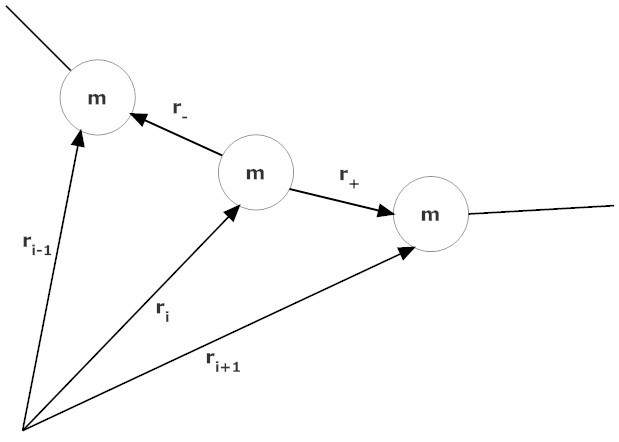}
	\caption{General diagram for a spring freely moving through space after being released from a rotating rotor.}
	\label{fig:generalspring}
\end{figure}

The spring is simulated as 100 point masses connected by short massless springs. The initial position of each point mass must be correctly determined so that it satisfies the quasi-static initial condition. Fig. \ref{fig:generalspring} shows the labeling of each mass.

Each vector locates the mass in all three dimensions. The spring forces are summed with air resistance and gravity and set equal to the mass times acceleration as follows.

\begin{equation}
+k(\vec{r}_{i+1}-\vec{r}_i)-k(\vec{r}_i-\vec{r}_{i-1})+\vec{A}_i-mg\hat{z}=m\ddot{\vec{r_i}}
\label{roteqn}
\end{equation}

The position vector $\vec{r}_i$ is the standard three dimensional Cartesian vector.
\begin{equation}
\vec{r}_i=x_i\hat{x}+y_i\hat{y}+z_i\hat{z}
\end{equation}

Air resistance $\vec{A}_i$ is directed opposite the velocity vector for each mass. The best results are obtained by setting air resistance proportional to the square of the mass velocity. \begin{equation}
\vec{A}_i=-bv_i^2\hat{v}_i
\end{equation}
The constant $b$ is adjusted for each spring to give the best results for all angular velocities examined.

Before release, the system is in its quasi-static state. It is moving in uniform circular motion with each mass maintaining the same radial distance from the center. Each mass will have a radial position close to the position calculated for the continuous case. So, in the simulation, the spring will start out in a perfectly straight horizontal position with the radial position of each mass determined by the solution for the continuous case. The stretched radial position of mass $i$ is initially given as $i\Delta{x}$
\begin{equation}
r_i=\frac{r_o}{cos(\phi)}\cos(\alpha (i\Delta{x}) +\phi)
\end{equation}

When the simulation starts, the spring will be pulled down by gravity and pushed back by air resistance. The system is allowed to evolve until all radial motion stops and all parts of the spring settles down into uniform circular motion. Resistive forces damp out transient motion. The first mass of the spring is attached to a central rotor which keeps the whole spring spinning at a set angular velocity, just like the experimental apparatus.

\subsection{Air Resistance and Angular Deflection}

As discussed above, the springs are observed to suffer a small amount of air resistance, causing the stretched spring to be deflected away from the angle of the rotor (see Fig \ref{fig:subfiguresa}). The simulation includes air resistance through a weak quadratic resistive term. 

\begin{equation}
A=-bv^2\hat{v}
\end{equation}

The initial transient motion of the simulated spring can be made to die out more quickly by applying a strong resistive force in the radial direction and vertical direction. In the quasi-static state there is no radial or vertical motion so air resistance in these directions can be set artificially high. In a sense, the masses are moving through ``syrup'' in the radial and vertical direction, allowing them to relax to their final positions. The air resistance tangential to the circular path of motion is its lower value that reflects the real effect of air resistance. When the spring is released, the ``syrup" like forces are removed and the spring moves under the influence of normal air resistance in all directions.

\begin{figure}[h!]
	\centering
	\includegraphics[height=2.0in]{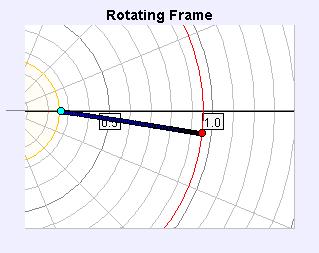}
	\caption{Simulated spring in the quasi-static state with drag before release. The inner circle is the path of the inner end of the spring, and the outer circle is the path of the outer end (or free end) of the spring.}
	\label{fig:simwithdrag}
\end{figure}

Once the transient oscillations die out, the spring returns to the quasi-static state but is no longer oriented radially outward from the center of the rotor. Fig. \ref{fig:simwithdrag} shows the simulated spring with an angular deflection of about 9 degrees due to air resistance. 

\subsection{Droop Angle}
The spring forms a fairly straight line that is bent downwards from the horizontal. The is referred to as the droop angle. As expected, the droop angle becomes smaller at larger angular velocities.

\subsection{True Rest Length}
There is a small improvement in the simulation if a ``True Rest Length" is introduced. The PASCO spring has a measured hook-to-hook rest length of 10.4 cm. However, the spring is under compression when at rest. It ``wants" to collapse to a smaller length but is prevented from doing so by the thickness of the spring's wire. The force in the spring is determined by how far the spring is displaced from its rest length, hence an estimate of the true rest length is needed. This corrected rest length was found by running the simulation for different rest lengths and checking against the results. For the PASCO spring a rest length of 8 cm gave the best results for the total spring extension while rotating.

\subsection{Non-Linear Spring }
The simulation method described above works well for a linear spring. That is a spring that reasonably obeys Hooke's Law for spring extensions seen in this experiment. The simulation for the PASCO spring works well treating the PASCO spring as linear. However, a Slinky behaves in a non-linear manner. 

\begin{figure}[h!]
	\centering
	\includegraphics[height=2.0in]{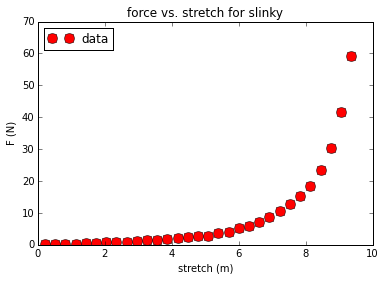}
	\caption{Force vs. stretch for the Slinky up to destruction.}
	\label{fig:slinkystretch}
\end{figure}

In our first version of the simulation, the simulated values and the data agreed for low angular velocities, but the Slinky extension failed at larger velocities. This was due to the Slinky's non-linear stretch at high angular velocities. The experimental Slinky was therefore tested to destruction. Fig. \ref{fig:slinkystretch} shows a graph of  force vs. stretch for the Slinky. When the non-linear stiffness of the Slinky is accounted for, the spring extension in the simulation agrees well with experimental data measured from video analysis.

The Slinky force as a function of stretch was fitted with a third order polynomial. This fit was sufficient for the extension that the Slinky experiences in this experiment. The function is given as,
\begin{equation}
	F=0.0133x^3-0.0256x^2+0.3718x
\end{equation}

where $x$ is the spring extension from rest in meters and $F$ is in newtons.

For a linear spring, the simulation starts with the spring rotating with the correct extension, mass distribution and angular velocity needed when there is no air resistance. With air resistance, the system takes a few simulation seconds to adjust and settle down into a quasi-static state. In the non-linear case, the spring starts out extended further than it should be, so additional time is needed to let the Slinky slowly relax into the correct extension and drag angle. This usually takes about 15 seconds of simulation time.

\section{Results}
\begin{figure}[h!]
	\begin{center}
		\subfigure[(one)]{%
			\label{fig:subfiguresa}
			\includegraphics[width=0.22\textwidth]{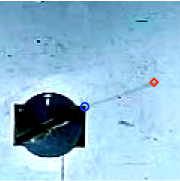}
		}%
		\subfigure[(two)]{%
			\label{fig:subfiguresb}
			\includegraphics[width=0.22\textwidth]{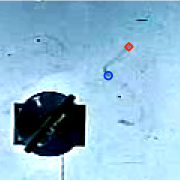}
		}\\ 
		\subfigure[(three)]{%
			\label{fig:subfiguresc}
			\includegraphics[width=0.22\textwidth]{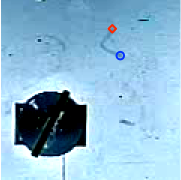}
		}%
		\subfigure[(four)]{%
			\label{fig:subfiguresd}
			\includegraphics[width=0.22\textwidth]{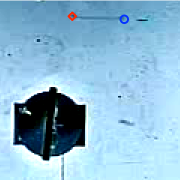}
		}%
	\end{center}
	\caption{%
		Four images of the spring after it is released. Note the ``chevron'' in the spring in the third figure.
	}%
	\label{fig:subfigures}
\end{figure}

\begin{figure}[h!]
	\begin{center}
		\subfigure[(one)]{%
			\label{fig:simsubfiguresa}
			\includegraphics[width=0.22\textwidth]{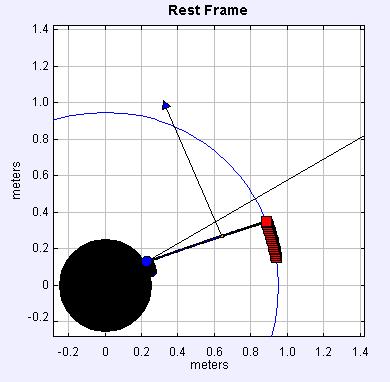}
		}%
		\subfigure[(two)]{%
			\label{fig:simsubfiguresb}
			\includegraphics[width=0.22\textwidth]{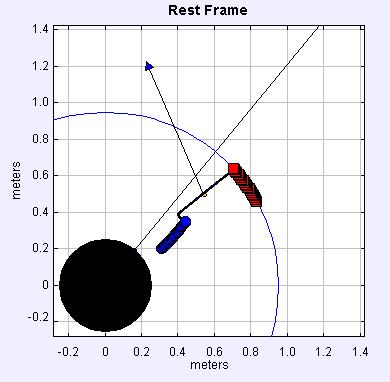}
		}\\ 
		\subfigure[(three)]{%
			\label{fig:simsubfiguresc}
			\includegraphics[width=0.22\textwidth]{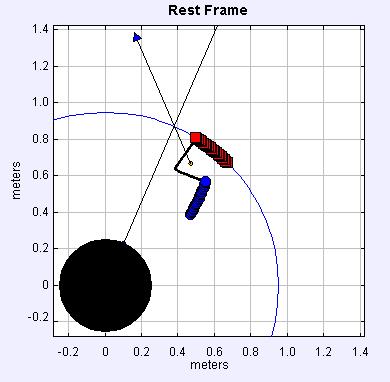}
		}%
		\subfigure[(four)]{%
			\label{fig:simsubfiguresd}
			\includegraphics[width=0.22\textwidth]{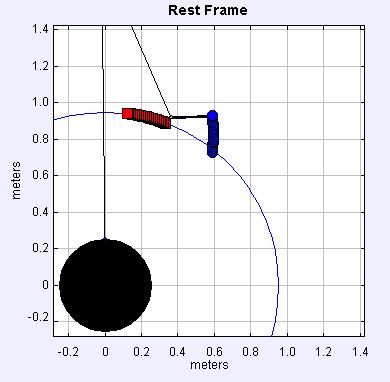}
		}%
	\end{center}
	\caption{%
		Four images of the spring after it is released. The heavy lines show the trajectories of the ends of the spring; the light connecting line is the string of point masses representing the spring. Note the ``chevron'' in the spring in the third figure.
	}%
	\label{fig:simsubfigures}
\end{figure}

Fig. \ref{fig:subfigures} shows images of the collapsing spring at various times after it is released. After release, the outer end continues on its original circular path and the inner end folds over to form a V-shape similar to a chevron. As a wave propagates from the inner end to the outer end of the spring, the collapsed portion of the spring (including the inner end) and the stretched portion of the spring (including the outer end) continue to make a chevron shape. As the wave propagates, the collapsed portion of the spring gets longer and the stretched portion of the spring gets shorter until the wave reaches the outer end and the spring is once again straight.  Fig. \ref{fig:simsubfigures} shows a similar set of frames from the simulation.

\begin{figure}[h!]
	\centering
	\includegraphics[height=2.8in]{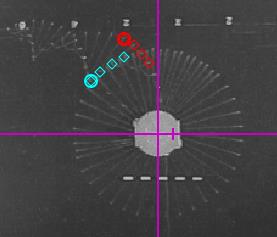}
	\caption{A strobographic photo of a 0.4 N/m spring rotating counterclockwise at 49 RPM prior to release. The outer end of the spring (blue) continues its circular motion for approximately 0.27 seconds after the inner end (red) is released, during which time it covers approximately 80$^\circ$ in angular displacement. }
	\label{fig:apparatus_release}
\end{figure}

When the wave reaches the outer end of the spring, it reflects and propagates back toward the inner end. This causes the motion of the spring to repeat (again forming a chevron shape), and each end of the spring follows a semi circular path with a radius equal to the quasi-static radius followed by the outer end of the spring. This is shown in the strobographic photo in Fig. \ref{fig:apparatus_release}.

\begin{figure}[h!]
	\centering
	\includegraphics[height=2.0in]{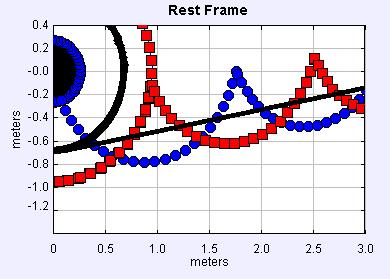}
	\caption{The paths of the inner and outer masses after release for one simulation. Note the semicircular paths.The straight line is the center of mass motion.}
	\label{fig:semicirc}
\end{figure}

Fig. \ref{fig:semicirc} Shows an example of the simulated semicircular motion from the simulation over a greater range. The long straight line is the center of mass motion after release. Once released, the system conserves angular momentum about its center of mass. As the wave propagates to the free end, angular momentum is being transfered to the released end. Finally, the free end is pulled from its original circular path and the angular momentum is transfered back causing it to move on a new semicircular path. This continues as the wave moves back and forth to each end of the spring with each end taking turns in forming the circular arc.

\begin{figure*}[h!]
\centering
\mbox{
\subfigure[PASCO spring]{
\includegraphics[width=0.5\textwidth]{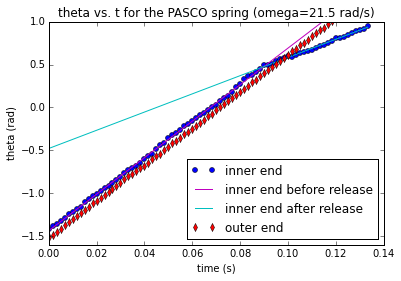}
}
\subfigure[Slinky]{
\includegraphics[width=0.5\textwidth]{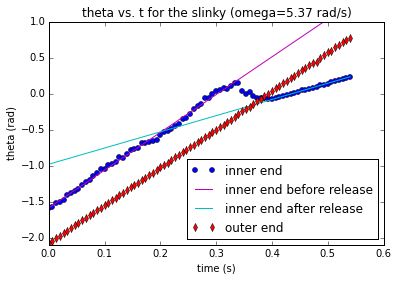}
}
}
\caption{$\theta(t)$ for the inner and outer ends of the PASCO spring and Slinky.}
\label{fig:thetavtdata}
\end{figure*}

A cursory view of the figures supports our hypothesis that after release, the free end of the spring would continue in uniform circular motion as before, until such time as a wave propagates from the inner end to the outer. From the data, we measured and graphed $\theta(t)$ for the inner end and outer end of the spring before and after release for both the PASCO spring and Slinky, with the origin at the axle of the rotor. See Fig. \ref{fig:thetavtdata}.

The slope of $\theta(t)$ for the outer end of the PASCO spring is the same before and after release and is the same as the slope of $\theta(t)$ for the inner end before release. After release, $\theta(t)$ for the inner end is linear showing that its angular velocity is constant, but with a smaller value than before release. The difference in the $\theta$-intercept for the inner end and outer end of the spring in Fig. \ref{fig:thetavtdata} is due to the outer end lagging behind the inner end as the spring rotates. This lag is caused by air resistance and is what we call the drag angle. The drag angle is visually apparent in the video frame in Fig. \ref{fig:subfiguresa} and in the simulation in Fig. \ref{fig:simwithdrag}.

The graph of $\theta(t)$ for the Slinky differs qualitatively from the PASCO spring when the inner end is released. Whereas the inner end of the PASCO spring experiences an abrupt change in $\omega(t)$ when it is released, the inner end of the Slinky experiences an angular acceleration during a longer time interval after it is released.  Thus, the jerk on the inner end of the PASCO spring is much larger (due to the smaller time interval) than the jerk on the Slinky.

\subsection{Measurable Quantities}
The quantities used to describe the motion of the spring fall into three categories: (1) the ``Static'' variables used as input parameters to the simulation; (2) the ``Quasi-static'' variables that describe the initial state of the spring on the rotating rotor; and (3) the ``Dynamic'' variables for the collapsing spring after it is released from the rotating rotor.

\subsubsection{Static Variables}
These are variables related to the spring and the system at rest.

\begin{itemize}
\setlength{\itemsep}{1pt}
\setlength{\parskip}{0pt}
\setlength{\parsep}{0pt}	
\item Spring Mass
\item Spring Constant
\item Rotor Radius
\item Spring Rest Length
\end{itemize}

\subsubsection{Quasi-static Variables}
These are measurements of the rotating spring before it is released.

\begin{itemize}
\setlength{\itemsep}{1pt}
\setlength{\parskip}{0pt}
\setlength{\parsep}{0pt}	
\item Spring Extension
\item Drag Angle
\item Droop Angle
\end{itemize}

The spring extension is the total length of the rotating spring before it is released. Air resistance causes the spring to lag behind the straight line defined by the radius of the rotor. The angle between the spring and the rotor is the drag angle. As the spring rotates, gravity pulls it down at an angle from the horizontal plane. This is called the droop angle.

\subsubsection{Dynamic Variables}

These are measurements of the system after the spring is released, while it is collapsing. There are many variables that can be measured, but the following two variables seem to be the most useful.

\begin{itemize}
\setlength{\itemsep}{1pt}
\setlength{\parskip}{0pt}
\setlength{\parsep}{0pt}	
\item Bend Angle
\item Collapse Time 
\end{itemize}

The bend angle is the angle formed by the chevron. The collapse time is the time interval for the wave to travel from the inner end of the spring to the outer end of the spring. 


%
%

\subsection{Measured Data, Simulated Values, and Graphs}

To test the veracity of the theoretical description of the quasi-static and dynamic state of the spring, the dependence of the quasi-static and dynamic variables on the angular velocity of the rotor were measured for two different springs: a standard PASCO  spring and a thin Slinky. The Slinky is a spring that stretches under its own weight, whereas the PASCO spring needs an extra external force to make it stretch. Five plots of interest are

\begin{itemize}
\setlength{\itemsep}{1pt}
\setlength{\parskip}{0pt}
\setlength{\parsep}{0pt}	
\item Collapse Time vs. Angular Velocity
\item Bend Angle vs. Angular Velocity
\item Spring Extension vs. Angular Velocity
\item Drag Angle vs. Angular Velocity
\item Droop Angle vs. Angular Velocity
\end{itemize}

Measurements of the quasi-static and dynamic variables for the PASCO spring (and Slinky) for various angular velocities are available on ComPADRE \cite{compadre} under the listing "Supplemental Data and Graphs". This file also contains tables of results from the computer simulations.

\subsubsection{PASCO Spring Data}

Measurements of the static variables for the PASCO spring at rest are listed in Table \ref{table:pascorestdata}. The true rest length is the rest length of the spring used in the simulation to give best results.

\begin{table}[h!]
	\begin{tabular}{|c|c|}
		\hline mass & 9.1 g \\ 
		\hline spring constant & 3.28 N/m \\ 
		\hline rotor radius & 25.3 cm \\ 
		\hline hook-to-hook rest length & 10.4 cm \\ 
		\hline coil-to-coil rest length & 7.2 cm \\ 
		\hline true rest length & 8.0 cm  \\
		\hline 
	\end{tabular}
	\caption{Data for static variables for the PASCO spring.}
	\label{table:pascorestdata}
\end{table}

\subsubsection{Simulated PASCO Spring Values}

The simulation was run for the PASCO spring for the same angular velocities measured with video analysis. Values determined from the simulation are available on ComPADRE \cite{compadre}.

\begin{figure*}[h!]
\centering
\mbox{
\subfigure[spring extension]{
\includegraphics[width=0.32\textwidth]{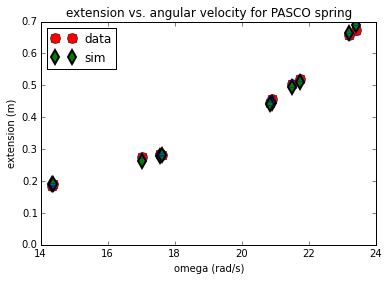}
}
\subfigure[drag angle]{
\includegraphics[width=0.32\textwidth]{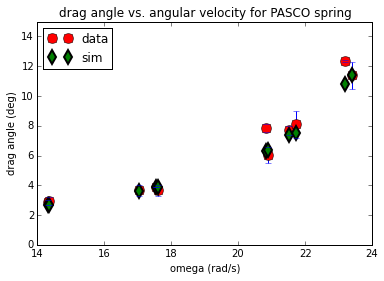}
}
\subfigure[droop angle]{
\includegraphics[width=0.32\textwidth]{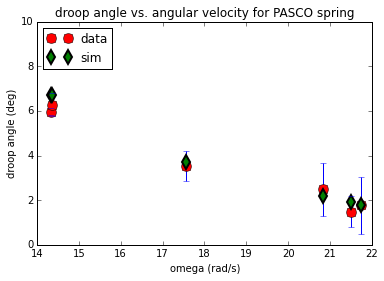}
}
}
\caption{Measured data and simulated values for quasi-static variables for the PASCO spring.}
\label{fig:pascosimdatafigqs}
\end{figure*}

Data from video analysis and values from the simulation for quasi-static variables are plotted in the three graphs shown in Fig. \ref{fig:pascosimdatafigqs}. 

\begin{figure*}[h!]
\centering
\mbox{
\subfigure[collapse time]{
\includegraphics[width=0.5\textwidth]{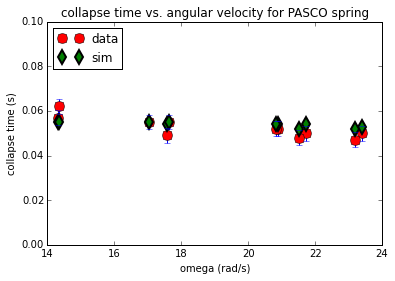}
}
\subfigure[bend angle]{
\includegraphics[width=0.5\textwidth]{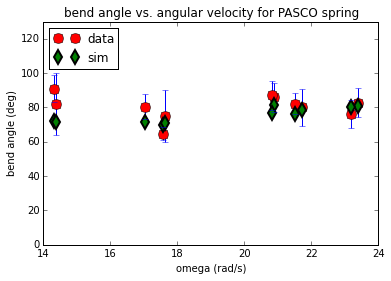}
}
}
\caption{Measured data and simulated values for dynamic variables for the PASCO Spring.}
\label{fig:pascosimdatafigdyn}
\end{figure*}

Data from video analysis and values from the simulation for dynamic variables are plotted in the two graphs shown in Fig. \ref{fig:pascosimdatafigdyn}. The simulation shows broad agreement with the experimental data for all dynamic variables.

\subsubsection{Slinky Data}

Measurements of the static variables for the Slinky at rest are listed in Table \ref{table:slinkdata}. 

\begin{table}[h!]
	\begin{tabular}{|c|c|}
		\hline mass & 28.485 g \\ 
		\hline spring constant & 0.38 N/m \\ 
		\hline rotor radius & 25.3 cm \\ 
		\hline coil-to-coil rest length & 3.3 cm \\ 
		\hline true rest length & 3.3 cm  \\
		\hline 
	\end{tabular}
	\caption{Data for static variables for the Slinky.}
	\label{table:slinkdata}
\end{table}  

Measurements of the quasi-static and dynamic variables for the Slinky for various angular velocities are available on ComPADRE \cite{compadre}. The Slinky has a larger mass and lower spring constant than the PASCO spring. This gives the Slinky a much greater extension than the PASCO spring. The angular velocities used for the Slinky are roughly one fourth the angular velocities used for the PASCO spring.


\subsubsection{Simulated Slinky Values}

The simulation was run for the Slinky for the same angular velocities measured with video analysis. The results of the simulation are available on ComPADRE  \cite{compadre}.

\begin{figure*}[h!]
\centering
\mbox{
\subfigure[spring extension]{
\includegraphics[width=0.32\textwidth]{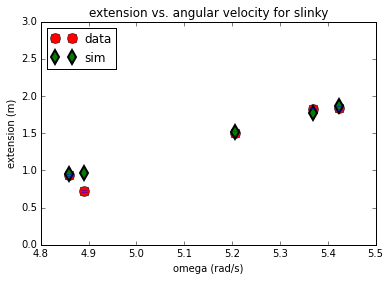}
}
\subfigure[drag angle]{
\includegraphics[width=0.32\textwidth]{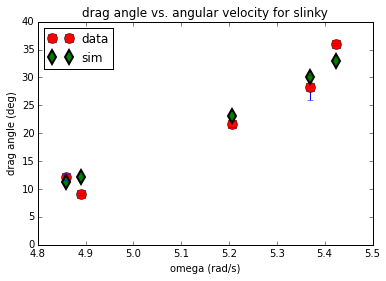}
}
\subfigure[droop angle]{
\includegraphics[width=0.32\textwidth]{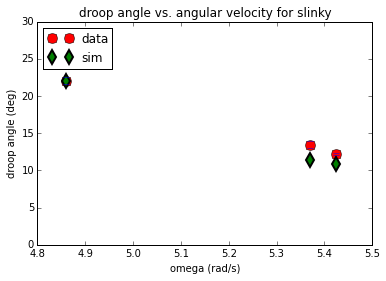}
}
}
\caption{Measured data and simulated values for quasi-static variables for the Slinky.}
\label{fig:slinkydataqs}
\end{figure*}

Data from video analysis and values from the simulation are plotted in the three graphs shown in Fig. \ref{fig:slinkydataqs} for the quasi-static variables.

\begin{figure*}[h!]
\centering
\mbox{
\subfigure[collapse time]{
\includegraphics[width=0.5\textwidth]{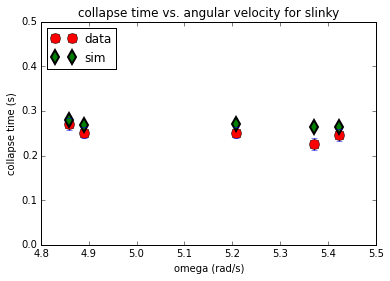}
}
\subfigure[bend angle]{
\includegraphics[width=0.5\textwidth]{data-kink-angle-omega-sim}
}
}
\caption{Measured data and simulated values for dynamic variables for the Slinky}
\label{fig:slinkydatadyn}
\end{figure*}

Data from video analysis and values from the simulation are plotted in the two graphs shown in Fig. \ref{fig:slinkydatadyn} for the dynamic variables.

\subsubsection{Collapse Time}
For a nonrotating spring under tension, the time for a wave to propagate down its length is independent of its stretch. Conceptually, the propagation time is longer for longer distances, but the propagation speed is likewise larger due to the greater tension required to stretch the spring. Taking the equation for wave velocity on a string,
\begin{equation}
\nu = \sqrt{\frac{T}{M/L}}
\end{equation}
where T is tension, M the spring's mass and L its length, Hooke's law is substituted for the tension. Assuming the unstretched length of the spring is small compared to the stretched length, the time for a wave to propagate down the spring is,
\begin{equation}
t=\frac{L}{\nu}=\frac{L}{\sqrt{\frac{(k)(\Delta L)}{M/L}}}\approx \frac{L}{\sqrt{\frac{(k)(L^2)}{M}}}=\sqrt{\frac{M}{k}}
\end{equation}
In the approximation, the time of collapse is independent of its length. In the rotating-spring situation described here, the tension varies along the length of the spring (maximum at the inner end, minimum at the free outer end) but the argument still holds in a piecewise and therefore global fashion down the length of a spring with varying tension. This is borne out by comparing $\sqrt{\frac{M}{k}}$ for the PASCO spring and the mini-Slinky (0.053 and 0.27 sec, respectively with the collapse times displayed in Fig. \ref{fig:slinkydatadyn} and Fig. \ref{fig:pascosimdatafigdyn}.


\section{Conclusion}
It has been demonstrated that when a vertical, hanging spring is released from rest, the bottom of the spring remains in equilibrium until a wave from the top end reaches the bottom. 

In this experiment, this property has been generalized to the case of a rotating spring. When rotating about a circle, the outer end of the spring continues on its circular path until a wave from the inner end reaches the outer end. 

Our computational model does a reasonable job of reproducing the data. Both the simulation and data show that upon release, there is an initial jerk on the inner end of the spring, but eventually the inner end will travel along a circle of constant radius and constant angular velocity as the spring is collapsing. In addition, the simulation and data agree that (1) the spring extension and drag angle increase with the angular velocity of the rotor; (2) the droop angle decreases with angular velocity of the rotor; (3) the collapse time and bend angle of the collapsing spring are independent of the angular velocity.

We feel this phenomenon serves as an easily-understood exemplar and introduction to the concept of locality. The original statement of locality is attributed to Issac Newton \cite{newton} ``��Tis unconceivable that inanimate brute matter should (without the mediation of something else which is not material) operate upon \& affect other matter without mutual contact." 

All forces are now described in terms of fields that preserve locality. From Einstein's space time field that allows for gravity waves to the quanta of force mediating fields. Classroom discussions of these topics, and even more esoteric topics such as quantum entanglement, may be more meaningful following an introduction to the concept of locality using the straightforward example of springs in circular motion presented here.

\section{Acknowledgements}

This research was initiated by Matthew Carnaghi, a student in Briana Fiser's first-year course for physics majors at High Point Unviersity called \emph{Research and Scientific Writing in Physics}. The idea for the project was first suggested by High Point University's Martin DeWitt.  Collaboration with Jeff Regester and Tom Dooling was a result of presentations at the annual North Carolina Section AAPT meetings (NCSAAPT). We acknowledge the benefit of undergraduate research in the freshman year and the valuable collaborations made possible by regional Sections of AAPT. 


\end{document}